%% file: main.tex
\documentclass[letterpaper,twocolumn,10pt]{article}
\usepackage{usenix-2020-09}

\usepackage{listings, listings-rust}
\usepackage{amsmath}
\usepackage{algorithm}
\usepackage[noend]{algorithmic}
\usepackage{mathtools}
\usepackage{listing}

\usepackage{tikz}
\usepackage{booktabs}   
\usepackage{subcaption} 

\usepackage[font=small,labelfont=bf]{caption}

\usepackage{amssymb}
\usepackage{amsmath}
\usepackage{amsthm}
\usepackage{comment}
\usepackage{pifont}
\usepackage{threeparttable}
\usepackage[normalem]{ulem}
\usepackage{tabu}
\usepackage{xspace}

\input{macros}

\begin{document}

\date{}

\title{\Large \bf Leveraging Large Language Models for Automated Proof Synthesis in Rust}

\author{
{\rm Jianan Yao}
\thanks{This work was done during an internship at Microsoft Research.}\\
Columbia University
\and
{\rm Ziqiao Zhou}\\
Microsoft Research
\and
{\rm Weiteng Chen}\\
Microsoft Research
\and
{\rm Weidong Cui}\\
Microsoft Research
} 

\maketitle
\pagestyle{plain}

\input{sections/0_abstract.tex}

\input{sections/1_intro.tex}
\input{sections/2_method}
\input{sections/3_eval}
\input{sections/4_conclusion}

\bibliographystyle{plain}
\bibliography{main}

\end{document}

%% file: macros.tex
\makeatletter
\newcommand{\eqnum}{\hfill\refstepcounter{equation}\textup{\tagform@{\theequation}}}
\makeatother

\definecolor{mygreen}{HTML}{467E7E}
\definecolor{mygray}{rgb}{0.5,0.5,0.5}
\definecolor{myblue}{HTML}{144B7D}
\definecolor{myorange}{HTML}{B25A00}
\definecolor{mymauve}{rgb}{0.88, 0.69, 1.0}
\lstdefinestyle{myC}{
  language=C,
  backgroundcolor=\color{white},
  basicstyle=\linespread{0.9}\ttfamily\small,
  breakatwhitespace=false,
  breaklines=true,
  commentstyle=[\bfseries\color{mygreen},    
  deletekeywords={...},
  escapeinside={<@}{@>},
  extendedchars=true,
  keepspaces=true,
  keywordstyle=\bfseries\color{myblue!70}, 
  otherkeywords={uint},     
  deletekeywords={get,angle, gamma, invariant},
  emph = { certiq_prove, match, end},
  emphstyle=\bfseries\color{myorange!70},
  showspaces=false, 
  showstringspaces=false,
  showtabs=false, 
  stringstyle=\color{mymauve},
  tabsize=1,
 morecomment=[f][\bfseries\color{mymauve!70}][0]{@},
 morecomment=[f][\bfseries\color{mygreen}][0]{//},
}

\definecolor{mred}{rgb}{.80,.12,.30}
\definecolor{grey}{rgb}{0.5,0.5,0.5}
\definecolor{purple2}{rgb}{.75,0,.85}
\definecolor{pistachio}{rgb}{0.58, 0.77, 0.45}
\definecolor{steelblue}{rgb}{.10,.40,.85}
\definecolor{forestgreen}{rgb}{.0,.60,.0}

\newcommand{\code}[1]{{\texttt{{\small #1}}}}

\newcommand{\cucomment}[1]{}

\renewcommand{\sout}[1]{}

%% file: sections/0_abstract.tex
\begin{abstract}
Formal verification can provably guarantee the correctness of critical
system software, but the high proof burden has long hindered its wide
adoption. Recently, Large Language Models (LLMs) have shown success in code
analysis and synthesis. In this paper, we present a combination of LLMs and
static analysis to synthesize invariants, assertions, and other proof
structures for a Rust-based formal verification framework called Verus. In a
few-shot setting, LLMs demonstrate impressive logical ability in generating
postconditions and loop invariants, especially when analyzing short code
snippets. However, LLMs lack the ability to retain and propagate context
information, a strength of traditional static analysis. Based on these
observations, we developed a prototype based on OpenAI's GPT-4 model.
Our prototype decomposes the
verification task into multiple smaller ones, iteratively queries GPT-4, and
combines its output with lightweight static analysis. We evaluated the
prototype with a developer in the automation loop on 20 vector-manipulating
programs. The results demonstrate that it significantly reduces human effort
in writing entry-level proof code.
\end{abstract}

%% file: sections/1_intro.tex
\section{Introduction}
\label{sec:intro}

Interactive formal verification addresses complex verification tasks that are
beyond the capabilities of push-button verification. However, using interactive
formal verification is more challenging because it demands significant manual
effort and specialized knowledge, especially when automatic verification fails.
Notably, the lines of code (LoC) required for verification can potentially
expand to as many as ten times the size of the original code
(\cite{ironfleet,veribetrkv}).

Inspired by recent advancements in Large Language Models (LLMs), we perceive
an opportunity to reduce manual efforts required for interactive formal
verification. We have developed a prototype that leverages OpenAI's GPT-4~\cite{gpt4} to
automate proof writing. This prototype specializes in programs that
operate on vectors. Leveraging GPT-4's capabilities in logical thinking and
code understanding, we aim to expedite the development of entry-level programs,
particularly those implementing well-known algorithms (e.g., sort, reverse).

Nevertheless, we have encountered two major challenges. The first challenge is
that GPT-4 does not strictly follow certain properties and proofs from earlier
contexts of a program. The second issue is that when an initial proof attempt
fails for a lengthy program, GPT-4 becomes overwhelmed by the multitude of error
messages, hindering its ability to improve the quality of proof.

To address these challenges, we divide a program into smaller segments,
and then utilize GPT-4 to generate the pre/post-conditions
for each segment. Subsequently, we ask GPT-4 to prove each segment
individually. This strategy allows GPT-4 to concentrate on a smaller segment
of the program per query. When GPT-4 generates a proof, our tool extends it
with a lightweight static analysis, which helps to propagate the properties
deduced from earlier sections of the program throughout the analysis process.
We evaluate our prototype on 20 vector-manipulating programs. For these
entry-level programs, our prototype tool reduces the LoC for proof
by over 80\%.

\section{Related work}
\label{sec:related}
There have long been efforts to automate various aspects of the verification pipeline, 
from automated invariant inference to tactic-based proof generation~\cite{coq_hammer, 
yang2019learning,czajka2020practical,first2023baldur}.
Invariant inference has been used to prove properties of loops~\cite{garg2013learning, 
garg2016learning, loopinvgen1, FreqHorn2, code2inv, cln2inv, pei2023can}, 
inductive algebraic data types~\cite{pldi21-beyond, HANOI}, 
and distributed protocols~\cite{ma2019i4, p-fol-ic3, swiss, induction_duality, DuoAI}. 
Among these lines of research, a growing number of methods are based on 
neural networks~\cite{yang2019learning, first2023baldur,  code2inv, cln2inv, pei2023can}, 
which has gained traction for verification tasks in recent years and has been shown to
better tackle the search space explosion problem
that has long hindered the scalability of traditional methods.
Different from those works, we apply an LLM to
synthesize invariants and intermediate assertions. Our work demonstrates that
future verification tools can be more efficient without sacrificing their
usability with the help of LLMs.

%% file: sections/2_method.tex
\section{Background}
We choose Verus~\cite{verus} as the base verification tool in our work.
Verus is a state-of-the-art verification tool for Rust that aggressively
prunes the SMT context to optimize solving time. Although it can verify large
and complicated systems more efficiently, it demands significantly more effort
to write proof code. To mitigate this, we consider several difficulties faced
by Verus developers. First, like many other verification languages, constructing
deductive invariants is hard due to the large search space. Second, since Verus
is very new to developers, it does not provide a large selection of reusable
proofs/lemmas or verified libraries. This requires developers to have an
understanding of logic and the ability to write proofs, even for basic
algorithms. Third, Verus encodes each module and even each loop independently to
facilitate fast solving. This optimization necessitates increased effort in
annotating the pre/post-conditions and invariants compared to other verification
languages (e.g., Dafny~\cite{dafny} or F*~\cite{fstar}).

\section{Methodology}
\label{sec:design}
\subsection{The need of auto-generated invariants}
\lstdefinestyle{codestyle}{
    basicstyle=\footnotesize,
    breaklines=true,
    frame=single,
    captionpos=b,
    numbers=left,
    xleftmargin=4.0ex,
    numberstyle=\tiny,
}
\begin{figure}[t]
\centering
\begin{lstlisting}[language=Rust,style=codestyle,escapechar=$]
fn reverse(v: &mut Vec<u64>)
ensures  $\label{line:postcondition_start}$
    v.len() == old(v).len(),
    forall|i:int| 0 <= i < old(v).len() ==> 
        v[i] == old(v)[old(v).len() - i -1] $\label{line:reverse_post_2}$
{
    let length = v.len();
    let mut n: usize = 0;
    while n < length / 2
    {
        let x = v[n];
        let y = v[length - 1 - n];
        v.set(n, y);
        v.set(length - 1 - n, x);
        n = n + 1;
    }
}
\end{lstlisting}
\caption{Function to reverse a vector. \code{ensures} specifies the
postcondition of the function. \code{old(v)} means the value of \code{v} before
the function executes.}
\label{fig:reverse}
\end{figure}

Consider a simple Rust program that reverses a vector, as shown in
Figure~\ref{fig:reverse}. The developer needs to verify two postconditions
specified at Lines~\ref{line:postcondition_start}-\ref{line:reverse_post_2}. The
first postcondition states that the reversed vector should maintain the same
length as the original vector, and the second postcondition states that the
$i$-th element of the reversed vector should be equal to the
$(\text{length}-i-1)$-th element in the original vector. These postconditions
define the correctness of the code. To prove the loop with
Verus~\cite{verus_loop_invariant}, the developer needs to add the following loop
invariants.

\begin{minipage}{.95\linewidth}
\begin{lstlisting}[language=Rust,style=codestyle,escapechar=$]
invariant
    0 <= n <= length / 2,
    v.len() == length,
    forall|i: int| 0 <= i < n ==> v[i] == old(v)[length - i - 1], 
    forall|i: int| length - n <= i < length ==> v[i] == old(v)[length - i - 1], 
    forall|i: int| n <= i < length - n ==> v[i] == old(v)[i],
\end{lstlisting}
\end{minipage}

Loop invariants define the conditions that remain true before and after
each iteration of the loop, and they should be inductive. The first invariant is
straightforward; it defines the conditions for the termination of the loop. The
second invariant is necessitated by Verus, as it performs separate verifications
for the loop and the other remaining parts of the program.

The third and fourth invariants specify the updates for any modified elements in
the vector, within the range $0 \leq i < n$ and $length - n \leq i < length$.
The final invariant asserts that every element that has not been updated retains
its initial value, ensuring that the invariants for updated elements are
inductive. The absence of any one of these invariants will lead to the failure
of establishing the inductive invariants.

\begin{figure*}[t]
    \centering
    \includegraphics[width=.75\linewidth]{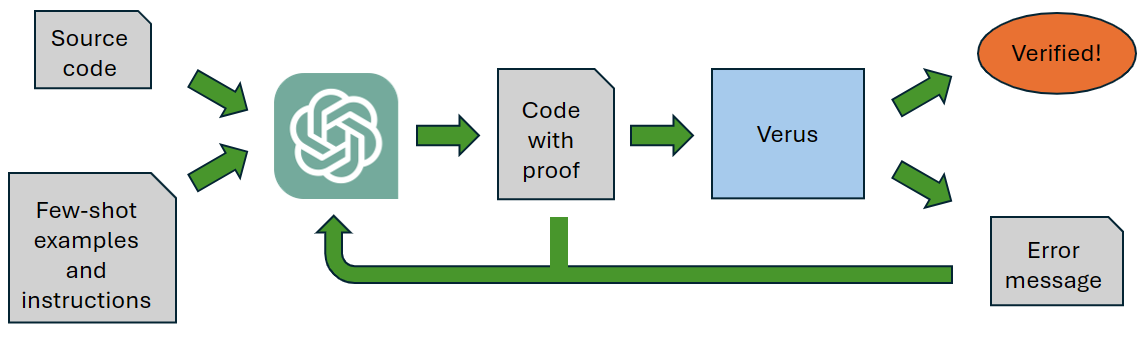}
    \caption{Basic workflow of our tool.}
    \label{fig:workflow}
\end{figure*}

To automatically fill these invariants (and potentially other proof structures),
we unleash the power of large language models in the workflow depicted in
Figure~\ref{fig:workflow}. Given the source code to be verified, we encode
it into a prompt with a few shot examples and send the prompt to GPT-4. Each
example is a pair of source code with to-be-proved properties (denoted as
\code{source\_code}) and verified code with human-provided proofs (denoted as
\code{code\_with\_proof}). When GPT-4 returns the code with proof, we
validate it by using Verus to verify it.

Most of the time, GPT-4 cannot solve the problem with a single query. If
verification with the generated proof fails, we follow a standard approach in
LLM chain-based solutions to integrate both the last response and
the error message to formulate a new prompt. This new query is then sent back to
GPT-4 for generating an improved result.

For the example in Figure~\ref{fig:reverse}, GPT-4 successfully generates
the first four invariants but misses the last one. In fact, human
developers often make the same mistake --- forgetting to specify things that do
not change. Verus then outputs three errors: the third and fourth invariants do
not hold at the end of the loop body, and the postcondition on
Line~\ref{line:reverse_post_2} does not hold. 

After incorporating the error message into the second query, GPT-4 returns
all 5 invariants, making the code verifiable by Verus. Ideally, if a human can
quickly adjust the proof based on the hints provided by the error messages, we
anticipate that GPT-4 can amend the proof in a similar manner.

\subsection{Task decomposition for large programs}
\label{sec:task_decomposition}
The basic solution described in the previous section is only effective for small programs. 
We have observed that GPT-4 does not perform well for relatively large programs.
It is not entirely surprising, given that we are asking GPT-4 to generate a complete proof for
the entire program, whereas human developers typically think in small
steps.

Consider the function in Figure~\ref{fig:sum}. A human developer would
initially prove the property of the vector following the first loop, asserting
that no element exceeds a value of two. Subsequently, they would shift their
focus to the second loop, analyzing its computation of a sum that does not
exceed twice the number of elements.

We can guide GPT-4 to think similarly by decomposing the large code task into
smaller ones. Given a code, we decompose it into smaller segments. For each
segment, we define two types of prompts. One is to let GPT-4 generate
the relationship between segments, where the post-condition of a segment must be
a pre-condition of the next segment. The other is to let GPT-4 generate the
proof per segment.

For the code in
Figure~\ref{fig:sum}, we can divide it at
Line~\ref{line:end_first_loop} into two segments and query GPT-4 for the postcondition of the
first segment. For example, GPT-4 gives the following postcondition.

\begin{lstlisting}[language=Rust,style=codestyle,escapechar=$]
i == N,
a.len() == N,
forall |k:int| 0 <= k < a.len() ==> a[k] <= 2,
\end{lstlisting}

With the postcondition, the verification of the original code is decomposed into
two smaller tasks, each concerning one of the two segments. For each segment, we
use the workflow depicted in Figure~\ref{fig:workflow} to complete the proof.
The three-line interface above will serve as the postcondition when verifying
the first segment and as the precondition when verifying the second segment.

\begin{figure}[t]
\centering
\begin{lstlisting}[language=Rust,style=codestyle,escapechar=$]
pub fn foo(a: &mut Vec<u32>, N: u32) 
    requires 
        old(a).len() == N,
        N <= 0x7FFF_FFFF,  $\label{line:upper_bound_N}$
{
    let mut i: usize = 0;
    while (i < N as usize)
    {
        if (a[i] > 2) {
            a.set(i, 2);
        } 
        i = i + 1;
    }  $\label{line:end_first_loop}$
    i = 0;
    let mut sum: u32 = 0;
    while (i < N as usize)
    {
        sum = sum + a[i];  $\label{line:overflow}$
        i = i + 1;
    }
    assert(sum <= 2 * N);
}
\end{lstlisting}
\caption{Verus function that sums over a vector after elements are capped at 2.
\code{requires} specifies the precondition of the function.}
\label{fig:sum}
\end{figure}

\subsection{Combining GPT-4 with static analysis and human}
Although GPT-4 can generate logical formulas based on code, including
complicated quantified invariants, they often overlook certain non-intuitive
simple invariants, much like beginner human developers.

For example, one might find the upper bound of \code{N} in Line~\ref{line:upper_bound_N}
confusing. However, this upper bound is crucial to ensure there is no
integer overflow on Line~\ref{line:overflow}. To verify the second loop, \code{N
<= 0x7FFF\_FFFF} must be included as an invariant. 

Finding such invariants is far more straightforward with static analysis. We have therefore implemented a
lightweight static analysis tool which, for each loop, provides lists of
variables read from and written to. For any precondition of the loop, if all
variables referenced in the precondition are not written to in the loop, and at
least one variable referenced is read from in the loop, the precondition will be
automatically propagated into the loop invariant, in addition to invariants
generated by GPT-4. For the second loop in Figure~\ref{fig:sum}, \code{N <=
0x7FFF\_FFFF} is added to the loop invariant.

To better guide GPT-4 query, we engage a developer only when necessary. If
GPT-4 returns a postcondition, the developer has the option to edit it. If GPT-4
fails to complete the proof within the maximum number of iterations, the developer is
prompted to complete the proof, building upon the partial proof provided by GPT-4.

\subsection{A running example}

To help understand how our tool works, we show each step for proving 
the first loop in Figure~\ref{fig:sum}.
\lstdefinestyle{userstyle}{
    basicstyle=\ttfamily\footnotesize,
    breaklines=true,
    frame=single,
    backgroundcolor=\color{blue!3},
    captionpos=b
}

\lstdefinestyle{gptstyle}{
    basicstyle=\ttfamily\footnotesize,
    breaklines=true,
    frame=single,
    backgroundcolor=\color{green!3},
    captionpos=b
}

\begin{lstlisting}[style=userstyle, caption={1st GPT-4 query}]
Given a program in Verus (a Rust-like language)
with its precondition, returns its postcondition.
Some examples and their postconditions are given
below:
${Examples}

Notes:
1. Do not use Rust primitives such as clone, filter, iter, sum, or count.
2. Do not use the conditional operator  "a ? b : c".
3. Each postcondition takes a separate line.
Begin! 
New Verus program:
${CodeBlock}

Response requirements:
You must respond using the following format and
ensure the response could be parsed by json.loads:
{
    "thoughts": "<explain the program and its precondition, then analyze how to write the postcondition>",
    "postcondition": "<postcondition>"
}
\end{lstlisting}
\begin{lstlisting}[style=gptstyle, caption={1st response. Our tool parses the response from GPT-4 and extracts the postconditions. }]
i == N,
forall |k:int| 0 <= k < N ==> a[k] <= 2
\end{lstlisting}
\begin{lstlisting}[style=userstyle, caption={2nd GPT-4 query, asking for a proof for this postcondition}]
Given a program in Verus (a Rust-like language),
add invariants and proofs to make it verifiable.
Some example programs and their verifiable
versions are given below:
...
Notes:
...
* Do not modify or remove any line in the given program. Only add lines.
Begin! 
New Verus program:
pub fn foo(a: &mut Vec<u32>, N: u32, i: usize) 
    requires 
        old(a).len() == N,
        N <= 0x7FFF_FFFF,
{
    let mut i: usize = i;
    assume(i == 0);
    ${first loop}
    proof {  //inline postcondition
        assert(i == N);
        assert(forall |k:int| 0 <= k < N ==> a[k] <= 2);
   }
}
------
Response requirements:
You must respond using the following format and
ensure the response could be parsed by json.loads:
{
    "thoughts": "<explain the pre and post conditions, and how to write invariants and proofs>",
    "verifiable program": "<program>"
}
\end{lstlisting}

\begin{lstlisting}[style=gptstyle, caption={2nd Response. GPT-4 outputs a program, which is the same as the input program except that it inserts the following invariants.
    }]
invariant
    0 <= i <= N,
    a.len() == N,
    forall |k:int| 0 <= k < i ==> a[k] <= 2,
\end{lstlisting}

In addition, our static analysis tool propagates the following invariants from the preconditions.
\begin{lstlisting}[style=userstyle]
a.len() == N,
N <= 0x7FFF_FFFF,
\end{lstlisting}

The loop invariant \texttt{N <= 0x7FFF\_FFFF} is then added to the GPT-generated invariants
(although it will not be necessary for this loop).
The program is then verified by Verus successfully.

%% file: sections/3_eval.tex
\section{Evaluation}
\label{sec:eval}

\subsection{Datasets}
We evaluated our tool on 20 vector-manipulating programs generated from the
Diffy~\cite{chakraborty2021diffy} benchmark. Specifically, we took 20
C programs from its \code{safe} category and translated them from C to Verus.
Then we manually checked the equivalence of the translation.

\subsection{Parameters}
We tested the verification capability of our tool, which is equipped with the
OpenAI GPT-4 (2023-03-15) model. Initially, we set the temperature of the GPT-4
model to 0. When GPT-4 returns a broken JSON format, the tool
increases the temperature to 0.5 and retries. If GPT-4 returns a
program that cannot be verified after invariant propagation, the tool feeds the
error message back and retries once. We utilized 3 prompt
templates: one for filling in the postcondition, one for completing the proof,
and one for fixing the proof. The static analysis is configured to divide a program
into segments around loops.

\subsection{Results}
\begin{table}
\centering
\begin{tabular}{l|r}
    \hline
    Total segments & 110 \\
    No proof needed & 55 \\
    GPT response verified directly & 18 \\
    Verified after invariant propagation & 17 \\
    Verified after error feedback & 2 \\
    Verified after both propagation and feedback & 1 \\
    Verified after human correction & 16 \\
    Unverified (buggy in Rust) & 1 \\
    \hline
\end{tabular}
\caption{Results on verifying the 20 programs by program segments.}
\label{tab:segment}
\end{table}

The 20 programs we tested were divided into 110 segments, resulting in a total
of 163 GPT-4 queries. Table~\ref{tab:segment} presents the results categorized by
program segments. Out of the 110 segments, 55 are loop-free and are directly
verified by Verus without requiring any annotations.
Of the remaining 55 segments, GPT-4 directly provides a correct proof for 18 of
them, while 20 segments receive a correct proof after invariant
propagation and/or error feedback. This showcases not only GPT-4's
inherent effectiveness but also the efficiency of the techniques we employ to
interact with it.

Table~\ref{tab:line_of_code} shows the results in terms of lines of code. When
starting from scratch, a human developer would require 334 lines of proof to
verify the 20 programs. In contrast, with our prototype tool, the user is tasked
with correcting only 55 lines, building upon the partial proof already provided
by the tool. This demonstrates the substantial reduction in human effort our tool
offers when verifying vector-manipulating programs with loops.

\begin{table}
\centering
\begin{tabular}{l|r}
    \hline
    Ground-truth proof & 334 \\
    Human corrections on syntax & 5 \\
    Human corrections on semantics & 49 \\
    Human corrections on both syntax and semantics & 1 \\
    \hline
\end{tabular}
\caption{Results on verifying the 20 programs by line of code.}
\label{tab:line_of_code}
\end{table}

\subsection{Improved results wth GPT-4 (2023-11-06)}

In our evaluation using the GPT-4 model dated 2023-03-15, only 3 out of 20
programs were fully automated (without human intervention). Additionally,
self-repair through error feedback was effective for only 2 segments. However,
after switching to the updated GPT-4 model (2023-11-06) and implementing two
additional attempts upon failure, 14 out of 20 programs required no human
intervention. With this enhanced automation, more than 20 segments could be self-repaired
via error message feedback. It demonstrates that our approach
naturally evolves alongside advancements in the LLM model.

%% file: sections/4_conclusion.tex
\section{Limitations and Lesson Learned}
\label{sec:lessons}
In this section, we share our experience and lessons learned when developing the tool. The first is
that GPT-4 works more effectively with shorter code inputs. When the code is long,
GPT-4 often forgets about invariants it writes for an earlier loop, and gets lost
in too many error messages when the proof is incorrect. Although the issue is
mitigated by task decomposition, as discussed in
Section~\ref{sec:task_decomposition}, the optimal strategy for
decomposition, especially with multiple functions, remains an area for research.

The second lesson is that code comments are appreciated by GPT-4.
We observed that GPT-4 sometimes forgets to specify the size of the vector in the
invariant (e.g., \code{v.len() == length}) for the reverse example in
Figure~\ref{fig:reverse}. By adding a comment after each such invariant in the
few-shot examples, GPT-4 is more likely to generate such an invariant for a new
program.


The third lesson is that GPT-4 is more adept at writing postconditions and
invariants than writing triggers and assertions for quantifier
instantiation~\cite{verus_trigger}, or writing nonlinear arithmetic
proof. Even in a zero-shot setting (i.e., when no example is provided in the
prompt), GPT-4 can produce meaningful postconditions and invariants, though not
in the valid Verus syntax. This indicates that GPT-4 has already
learned these concepts in its training data. But triggers and assertions for
quantifier instantiation are specific to annotation-based verification
languages, and proofs for nonlinear arithmetic are particularly specific to
Verus. Determining how to efficiently teach LLMs these new ways of reasoning
within a limited prompt size is an ongoing challenge. It is possible to solve
this problem by fine-tuning.

Our current tool is still an early prototype. The implementation specifically
targets single-function vector-manipulating programs in Verus. We anticipate its
capabilities would significantly expand by supporting more complex data types,
such as \code{Set}, \code{Map}, and user-defined datatypes. Another avenue for
enhancement would be to support cross-function verification and to leverage
existing lemmas in proofs.

\section{Conclusion}
In this paper, we presented an approach to use GPT-4 to generate proofs for Rust
programs that can be verified by Verus. We developed a prototype and evaluated
it on 20 vector-manipulating programs. Our evaluation shows that our prototype
can significantly reduce the human effort in writing proofs for entry-level programs.
Our work demonstrates the potential of leveraging LLMs to automate proof generation
for program verification.


\section{Acknowledgement}
We thank Chris Hawblitzel and Jacob R. Lorch for helpful suggestions on using Verus.